%% file: main.tex
\pdfoutput=1

\documentclass[11pt]{article}

\usepackage[final]{acl}

\usepackage{times}
\usepackage{latexsym}

\usepackage[T1]{fontenc}

\usepackage[utf8]{inputenc}

\usepackage{microtype}

\usepackage{inconsolata}

\usepackage{graphicx}
\usepackage{amsmath}

\usepackage{times}
\usepackage{epsfig}
\usepackage{graphicx}
\usepackage{amsmath}
\usepackage{amssymb}

\usepackage{wrapfig}
\usepackage{blindtext}
\usepackage{multirow}

\usepackage{booktabs}
\usepackage{tabularx}
\usepackage{xcolor,colortbl}
\usepackage{tabularray}
\usepackage{subcaption}
\usepackage{float}
\usepackage{stfloats}
\usepackage[most]{tcolorbox}
\usepackage[table,xcdraw]{xcolor}
\usepackage{array}
\usepackage{makecell}

\input{commands/mycommands}

\newcommand{\myalg}{\texttt{ContextualRetriever}}
\usepackage[hang,flushmargin]{footmisc}
\definecolor{pastelgreen}{rgb}{0.82, 0.94, 0.75}
\title{Learning Contextual Retrieval for Robust Conversational Search}

\author{Seunghan Yang$^1$ \quad Juntae Lee$^1$ \quad Jihwan Bang$^1$
\\ {\bf Kyuhong Shim$^{2}$\footnotemark[1] \quad Minsoo Kim$^{1}$\footnotemark[1] \quad Simyung Chang$^{1}$\footnotemark[1]}\\
{$^1$Qualcomm AI Research\footnotemark[2]} \quad {$^2$Sungkyunkwan University} \\
{\texttt {\small\{seunghan, juntlee, jihwbang\}@qti.qualcomm.com}} \quad {\texttt {\small khshim@skku.edu}}}

\begin{document}
\maketitle
\renewcommand{\thefootnote}{\fnsymbol{footnote}} 
\footnotetext[1]{Work completed while employed at Qualcomm AI Research.}
\footnotetext[2]{Qualcomm AI Research is an initiative of Qualcomm Technologies, Inc.}

\begin{abstract}
Effective conversational search demands a deep understanding of user intent across multiple dialogue turns. Users frequently use abbreviations and shift topics in the middle of conversations, posing challenges for conventional retrievers. While query rewriting techniques improve clarity, they often incur significant computational cost due to additional autoregressive steps. Moreover, although LLM-based retrievers demonstrate strong performance, they are not explicitly optimized to track user intent in multi-turn settings, often failing under topic drift or contextual ambiguity.
To address these limitations, we propose \myalg{}, a novel LLM-based retriever that directly incorporates conversational context into the retrieval process. Our approach introduces: (1) a context-aware embedding mechanism that highlights the current query within the dialogue history; (2) intent-guided supervision based on high-quality rewritten queries; and (3) a training strategy that preserves the generative capabilities of the base LLM. Extensive evaluations across multiple conversational search benchmarks demonstrate that \myalg{} significantly outperforms existing methods while incurring no additional inference overhead.
\end{abstract}

\input{contents/01_intro}
\input{contents/05_related_work}
\input{contents/03_method}
\input{contents/04_exp}
\input{contents/06_conclusion}

\bibliography{custom}

\newpage
\input{contents/07_Appendix}

\end{document}

%% file: commands/mycommands.tex
\usepackage{geometry}
\usepackage{booktabs} 
\usepackage{bbm}
\usepackage{mathtools}
\usepackage{nccmath}
\usepackage{setspace}

\usepackage{caption}
\usepackage{subcaption}

\usepackage[linesnumbered,ruled,vlined]{algorithm2e}

\SetKwInput{KwInput}{Input}                
\SetKwInput{KwOutput}{Output}              

\SetCommentSty{mycommfont}

\SetAlCapSty{algcapsty}

\usepackage[T1]{fontenc}
\usepackage{wrapfig,lipsum,booktabs}

\usepackage{soul}
\usepackage{dsfont}
\usepackage{enumerate}
\usepackage{enumitem}

\usepackage{amsmath}
\usepackage{amsfonts}
\usepackage{bbm}
\usepackage{dsfont}
\usepackage[Symbol]{upgreek}
\usepackage{lscape}
\usepackage{caption}
\usepackage{balance}
\usepackage{xspace}
\usepackage{float}
\usepackage{kotex}

\usepackage{wasysym}
\usepackage{xcolor}
\usepackage{multirow}
\usepackage{array, boldline, rotating}

\usepackage{amssymb}
\usepackage{pifont}
\renewcommand*\eqref[1]{(\ref{#1})}

\definecolor{LightCyan}{rgb}{0.88,1,1}
\definecolor{Blue}{rgb}{0, 0.5, 1}
\definecolor{Green}{rgb}{0.0, 0.8, 0.0 }
\definecolor{Red}{rgb}{0.95, 0.55, 0.6}
\definecolor{Skyblue}{rgb}{0.6, 0.6, 0.95 }

\NewDocumentEnvironment{suptitle}{ +b }{
    \twocolumn[{#1}]%
}{}

\NewDocumentCommand{\supptitle}{s}{
\begin{suptitle}
        \centering
        \rule{\textwidth}{0.03cm}\\[0.1cm]
        Appendix\\[0.2cm]
        {\Large 
            \textbf{Learning Contextual Retrieval for Robust Conversational Search}
        }\\
        \rule{\textwidth}{0.03cm}\\[0.2cm]
\end{suptitle}}

\NewDocumentCommand{\summarytitle}{s}{
\begin{suptitle}
        \centering
        \rule{\textwidth}{0.03cm}\\[0.1cm]
        {\Large 
            \textbf{Learning Contextual Retrieval for Robust Conversational Search}
        }\\
        \rule{\textwidth}{0.03cm}\\[0.2cm]
\end{suptitle}}

%% file: contents/01_intro.tex
\section{Introduction} 
\begin{figure}[t]
    \centering
    \includegraphics[width=1.0\linewidth]{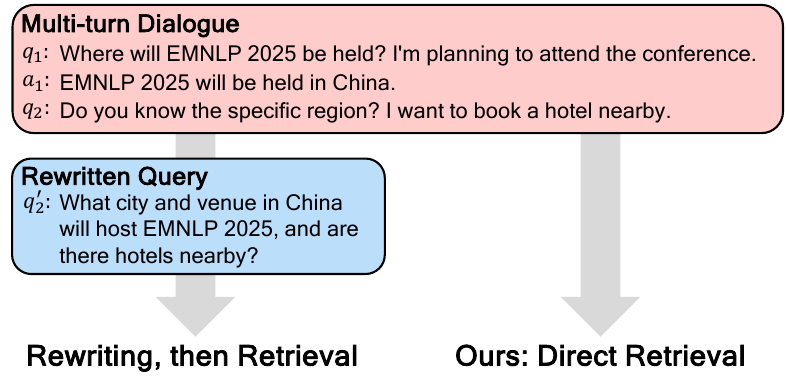}
\vfill
\centering
\resizebox{\linewidth}{!}{
\begin{tabular}{l|c|c}
\toprule
Method & {Infer. time\textsuperscript{↓}} & {Performance\textsuperscript{↑}} \\ \midrule
Naive unified LLM retriever  & 80.3 ms & 64.2\% \\
\;\;\;+ Rewriting (on its own) & 1100.5 ms & 77.2\% \\ \midrule
\rowcolor{pastelgreen}
\textbf{Ours (internalized contextualizing)} & \textbf{80.5 ms} & \textbf{91.9\%} \\ \bottomrule
\end{tabular}
}
\caption{\textbf{Potential of LLM-based retriever to contextualize the query in conversational search.} Even a naive unified LLM retriever can rewrite the query and generate embeddings on its own, improving retrieval performance from 64.2\% to 77.2\%. Our proposed method, \myalg{}, achieves 91.9\% by better leveraging the contextual understanding capabilities of LLMs, without additional inference overhead.}
\label{fig:intro}
\vspace{-1mm}
\end{figure}

The rapid advancement of chatbots has significantly increased demand for conversational search engines~\cite{ConvSearch1, ConvSearch3}. These systems must accurately retrieve information from large document collections to provide reliable, factual responses. Traditional search engines primarily handle single-turn queries and struggle in multi-turn conversational contexts, particularly when users heavily rely on abbreviated or context-dependent queries, as exemplified by $q_2$ in Figure~\ref{fig:intro}. Thus, effective contextualization, which involves understanding user intent throughout the conversation, is crucial for accurate retrieval.

A common strategy is query rewriting~\cite{T5QR, ConvGQR}, which reformulates abbreviated or ambiguous user queries into fully specified ones by integrating conversational context (see Figure~\ref{fig:intro}, blue box). While it improves clarity, it requires additional rewriting models, increasing inference time and computational overhead.

Recent approaches aim to build retrievers by directly fine-tuning Large Language Models (LLMs), leveraging their inherent language understanding capabilities~\cite{Mistral, Qwen, Llama}. LLM-based retrievers such as SFR Embedding~\cite{SFR} and NV-Embed~\cite{NV1, NV2} apply contrastive learning to pretrained LLMs, optimizing them specifically for retrieval tasks.
Unified retrievers, such as GritLM~\cite{GritLM} and OneGen~\cite{OneGen}, handle generation and retrieval tasks via multi-task learning within a single model. However, these methods primarily target single-turn queries and have not fully utilized LLMs' ability to model multi-turn conversational context.
To incorporate conversational history, ChatRetriever~\cite{ChatRetriever} compresses prior turns into a limited number of special tokens. However, this compression strategy quickly saturates, yielding only marginal improvements as more tokens are added. This suggests a key limitation of current approaches: they fail to deeply embed rich conversational history into the retrieval representation itself.

Although prior work has made meaningful progress, few studies have fully leveraged the language understanding capabilities of LLMs to contextualize conversational queries.
We hypothesize that LLM-based retrievers possess strong potential for modeling user intent across turns, but this ability remains underutilized in current designs.
To validate this hypothesis, we examine GritLM, a unified LLM retriever trained for both generation and retrieval. We compare its retrieval performance when using embeddings derived from its own rewritten queries versus embeddings obtained directly from the original user queries. As shown in Figure~\ref{fig:intro}, rewritten queries yield significantly better retrieval accuracy, suggesting that the model captures user intent and context well during the rewriting step. However, this contextual understanding is not effectively reflected in the retrieval embeddings generated from the original queries, indicating that the model's ability to embed conversational context remains underexploited.
This insight motivates our work: we aim to directly encode user intent and dialogue context into the retrieval representation itself, enabling LLM-based retrievers to fully capitalize on their inherent contextual understanding ability.

To address this limitation, we introduce \textbf{\myalg{}}, a novel approach designed to better harness LLMs for retrieval in multi-turn conversations. \myalg{} comprises three core components:
First, it employs a context-aware embedding mechanism that emphasizes the current query while encoding the full dialogue. Retrieval embeddings are computed solely from the current query segment, maintaining focus on the immediate information need while grounding it in broader conversational context.
Second, it leverages intent-guided supervision by aligning model-generated embeddings with those derived from high-quality rewritten queries. These rewritten queries clarify user intent, allowing the model to learn intent-aware representations without requiring an explicit rewriting step at inference time.
Third, it incorporates generation loss during training to preserve the LLM’s intrinsic language understanding capabilities. This allows the model to retain its general linguistic competence, which is essential for interpreting ambiguous or context-dependent queries.

We evaluate \myalg{} on four standard conversational search benchmarks: TopiOCQA~\cite{TopiOCQA}, QReCC~\cite{QReCC}, TREC-CAsT~\cite{TREC19, TREC20}, and ORConvQA~\cite{ORConvQA}. Our method consistently outperforms strong baselines, demonstrating that embedding user intent directly into the retrieval space substantially improves multi-turn conversational search without introducing additional inference overhead.

%% file: contents/05_related_work.tex
\section{Related Works}
\subsection{Dense Retrieval}
Information retrieval has evolved from traditional lexical matching methods such as BM25 and TF-IDF~\cite{BM25, TF-IDF} to dense retrieval approaches~\cite{DPR, ColBERT}. Dense retrievers encode queries and passages into vector embeddings and perform retrieval based on their similarity.

Early dense retrievers built on BERT~\cite{BERT} leverage its contextual representation power~\cite{BGE, MiniLM}. More recent approaches~\cite{SFR, NV1, NV2, BGE-ICL} utilize larger pre-trained LLMs to take advantage of superior language understanding. However, these models are typically trained on isolated query-passage pairs, which limits their ability to understand conversational context. Unified LLM retrievers such as GritLM~\cite{GritLM} and OneGen~\cite{OneGen} attempt to combine generation and retrieval in a single model for efficiency, but they fall short in embedding rich conversational context during retrieval.

\begin{figure*}[t]
    \centering
    \includegraphics[width=1.0\linewidth]{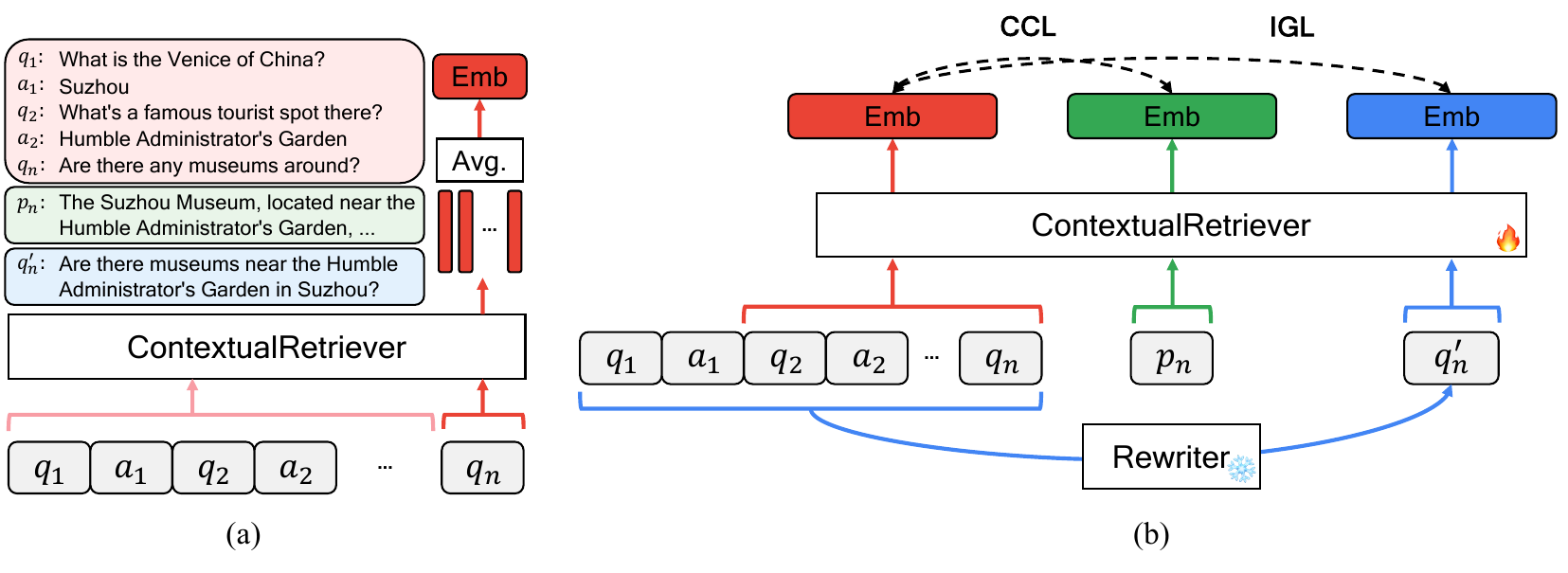}
    \vspace{-9mm}
    \caption{\textbf{Overview of our \myalg{}.} (a) \myalg{} processes the entire conversation history including the current query, but extracts retrieval embeddings specifically focused on the current query. (b) \myalg{} with the frozen rewriter is trained to align the query embeddings with both relevant passage embeddings and rewritten query embeddings for effective context understanding through Conversational Contrastive Learning (CCL) and User Intent-Guided Learning (IGL).}
    \label{fig:overview}
    \vspace{-4mm}
\end{figure*}

\subsection{Conversational Search}
Most dense retrievers are trained on single-turn settings with clearly stated information needs. In contrast, conversational search introduces challenges such as ambiguity and context dependence. Query rewriting approaches~\cite{T5QR, ConvGQR} reformulate conversational queries into self-contained forms, but they incur significant computational overhead due to their reliance on a separate rewriting model.

Recent efforts have aimed to integrate conversational context directly into retrievers. CQE~\cite{CQE} and ConvAUG~\cite{CONVAUG} generate context-aware embeddings via contrastive learning and data augmentation, respectively, but they do not explicitly model contextual ability. ConvDR~\cite{ConvDR} and DiSCo~\cite{DiSCo} leverage knowledge distillation from rewritten queries to embed context, though their effectiveness heavily depends on the quality of the rewriting model. Shortcut Dependency~\cite{ShortCut} improves retrieval robustness by mitigating shortcut learning from topical cues, but it lacks deeper semantic modeling of dialogue context. ChatRetriever~\cite{ChatRetriever} leverages LLMs by encoding dialogue history into special tokens, but its performance quickly saturates, yielding marginal gains as context length increases.
In contrast, our approach leverages the contextual capabilities of LLM by combining query rewriting–based supervision with generation-based training. This allows the model to encode both explicit intent signals and implicit contextual information without relying on external rewriting modules.

%% file: contents/03_method.tex
\section{Method}
\subsection{Task Definition}
Conversational search~\cite{ConvSearch3} aims to retrieve relevant passages from a collection $P = \{p_1, \ldots, p_m\}$ for each query in multi-turn dialogues.
At the $n$-th conversation turn, the goal is to retrieve top-k passages for the current query $q_n$ by leveraging the conversation history $\{q_i, a_i\}_{i=1}^{n-1}$, where $q_i$ and $a_i$ denote the query and response at the $i$-th turn, respectively.
The retriever $R(\cdot)$ encodes both passages and queries into a shared embedding space. Each passage is pre-encoded offline, while the current query, together with its conversation history, is encoded during inference. Retrieval is performed by computing the cosine similarity between the query and passage embeddings.

\subsection{Construction of Training Set}
\label{sec:sampling}
We introduce a \textit{dynamic dialogue history sampling strategy}, creating varied training instances from conversation histories given a target query-passage pair $(q_n, p_n)$.  
Specifically, we randomly select a starting point $i$ $(i\!<\!n)$ and include all subsequent queries and responses $[q_i, a_i, \ldots, q_n]$ to form training pairs with the relevant passage $p_n$. This approach (1) augments the training data, (2) exposes the model to diverse context lengths, and (3) improves robustness to varying conversational histories. Our experiments confirm that this strategy significantly boosts the model’s capacity to incorporate conversational context and generate high-quality retrieval embeddings (Table~\ref{ablation3}).

\subsection{Retrieval Embedding Extraction}
\label{sec:extraction}
We design our retriever based on a decoder-only LLM architecture, following the previous LLM-based retrievers. As illustrated in Fig.~\ref{fig:overview}(a), while our model takes the entire conversation history as input, it selectively extracts embeddings only from the tokens corresponding to the current query. This selective extraction enables the model to effectively utilize the conversational history as contextual cues while ensuring the retrieval remains strongly aligned with the intent of the current query. This approach mitigates the risk of excessively prioritizing prior conversation context, which could otherwise hinder retrieval accuracy by overshadowing the immediate query intent.
The retrieval embedding for the current query is computed by average pooling the sequence embeddings of the last $m$ elements:
\begin{equation}
    e_{q_n} = \mathrm{AvgPool}(\{R([q_i, a_i, \ldots, q_n])_j\}^N_{j=N-m+1}),
    \label{eq1}
\end{equation}
 where $R([q_i, a_i, \ldots, q_n])_j$ represents the embedding for the $j$-th token within the sequence. Here, $m$ and $N$ denote the token length of the current query and input conversation, respectively.

\subsection{Training for Conversational Search}
\subsubsection{Conversational Contrastive Learning}
We optimize the retriever to distinguish relevant from irrelevant passages using a contrastive learning objective applied to our constructed conversational dataset:
\begin{equation}
    L_{CCL} = -\log\frac{f(q_n, p_n)}{f(q_n, p_n) + \sum_{p_k\in P_n^{-}}f(q_n, p_k)},
\end{equation}
where $f(q_n, p_n) = \exp((e_{q_n} \cdot e_{p_n}) / \tau)$ is a similarity function with temperature $\tau$, and $P_n^{-}$ denotes a set of negative passages for query $q_n$.
This contrastive framework serves two key purposes. First, it shapes the embedding space by pulling relevant query-passage pairs closer and pushing negatives apart. Second, because query embeddings are computed with the full dialogue context, $L_{CCL}$ implicitly encourages the model to encode contextual information that improves retrieval performance.

\subsubsection{User Intent-Guided Learning}
To further enhance our retriever's ability to capture user intent, we propose an intent-guided learning approach that leverages signals from query rewriting.
Our method employs LLMs as a query rewriter $QR(\cdot)$ to generate contextually explicit queries through carefully designed prompts (See Appendix~\ref{query_rewriting_prompts}). The rewriter transforms abbreviated queries into self-contained formats by incorporating relevant context from previous interactions.
We introduce an embedding alignment loss that bridges the gap between the embeddings of the original and rewritten queries:
\begin{equation}
     L_{IGL} = \left\| e_{q_n} - e_{q'_n}\right\|^2_2,
\label{IGL}
\end{equation}
where $q'_n = QR([q_1, a_1, \ldots, q_n])$ represents the rewritten query.
While conversational contrastive learning optimizes query-passage relationships, intent-guided learning focuses on aligning query representations with their explicit, context-aware counterparts. As shown in Fig.~\ref{fig:overview}(b), these learning objectives work together to ensure our model leverages comprehensive intent understanding to achieve effective retrieval performance.

\subsubsection{Preserving LLM Capabilities}
To maintain the rich language understanding capabilities of the base LLM while optimizing for retrieval performance, we introduce a generation-based regularization technique that shares the same computational path with retrieval. Specifically, we employ a next-token prediction loss that encourages the model to preserve its inherent ability to generate contextually appropriate responses:
\begin{equation}
    L_{G} = -{\log}P(R(a_n)|R([q_i, a_i, ..., q_n, p_n])),
\end{equation}
where the model predicts the next response $a_n$ given the conversation and relevant passage.

\subsubsection{Final Training Objective}
Our complete training objective is:
\begin{equation}
    L = (1-\lambda_{G}) (L_{CCL} + \lambda_{IGL}L_{IGL}) + \lambda_{G}L_G,
\label{final_loss}
\end{equation}
where $\lambda_{IGL}$ and $\lambda_{G}$ control the balance among the loss components.
We refer to the final retriever trained with this complete objective as \textbf{\myalg{}}.

%% file: contents/04_exp.tex
\section{Experiment}
\subsection{Experimental Setup}
\noindent\textbf{Datasets.}
We evaluate our approach on four widely-used conversational search datasets: TopiOCQA~\cite{TopiOCQA}, QReCC~\cite{QReCC}, TREC-CAsT~\cite{TREC19, TREC20}, and ORConvQA~\cite{ORConvQA}. All datasets feature multi-turn conversational queries, containing both current queries and conversation history.
TopiOCQA contains frequent topic shifts within a conversation, requiring systems to determine whether to maintain or discard prior context.
QReCC and ORConvQA are relatively topic-consistent, where the primary challenge is resolving context-dependent expressions such as pronouns and ellipses by referencing prior turns.
TREC-CAsT 2019 and 2020 feature evolving user information needs within a controlled experimental setup. Conversations average around 9 to 10 turns and are manually curated to ensure coherence and diversity. We train our \myalg{} on TopiOCQA’s training split and evaluate it on all four datasets: in-domain (TopiOCQA test set) and out-of-domain (QReCC dev set, TREC-CAsT test sets, ORConvQA test set). Statistics of each dataset are reported in Appendix~\ref{data_stat_section}.

\noindent\textbf{Evaluation.}
We employ standard information retrieval metrics to evaluate retrieval effectiveness, including Mean Reciprocal Rank (MRR), normalized Discounted Cumulative Gain at rank 3 (nDCG@3), and Hit Rate at rank k (Hit@k).
MRR captures how early the first relevant document appears in the ranking.
nDCG@3 evaluates both the presence and ranking quality of relevant documents within the top-3 results.
Hit@k denotes the proportion of queries where at least one relevant document is retrieved within the top-k candidates.

\input{tables/ablation1}

\noindent\textbf{Implementation details.}
We apply LoRA-based fine-tuning of our method to three different retrievers: BGE-large~\cite{BGE}, SFR Embedding~\cite{SFR}, and GritLM~\cite{GritLM}. As shown in Table~\ref{ablation1}, our method consistently improves performance across all models, with particularly strong gains when applied to GritLM. We attribute this compatibility to GritLM's joint training objective for generation and retrieval, which aligns naturally with our generation-preserving learning objective. Given this synergy, we select GritLM as the base retriever for our main experiments.
We use LoRA with the following hyperparameters: 1 training epoch, batch size of 24, learning rate of 1e-4, LoRA rank of 16, and Adam optimizer. The weights for intent-guided learning ($\lambda_\mathrm{IGL}$) and generation loss ($\lambda_\mathrm{G}$) are set to 1.0 and 0.2, respectively.

\subsection{Baselines}
\noindent\textbf{Query rewriter.} We consider three query rewriting approaches: T5QR~\cite{T5QR}, GritLM~\cite{GritLM}, and GPT-4-Turbo~\cite{GPT4}. T5QR is a dedicated query rewriting model fine-tuned from the T5-base architecture~\cite{t5base} using the TopiOCQA training set. In contrast, GritLM and GPT-4-Turbo are general-purpose language models that we leverage for prompt-based query rewriting, following recent trends in LLM-driven conversational rewriting~\cite{ConversatioanlReWrite} (see Appendix~\ref{query_rewriting_prompts}).

\noindent\textbf{Retriever.}
We consider two types of dense retrievers. First, we evaluate BERT-based models, including MiniLM~\cite{MiniLM} and BGE-large~\cite{BGE}, which are efficient and strong general-purpose retrievers~\cite{MTEB}. Second, we evaluate LLM-based retrievers, including SFR Embedding~\cite{SFR}, a retrieval-specialized Mistral-7B model; GritLM~\cite{GritLM}, a unified retriever-generator also based on Mistral-7B; and ChatRetriever~\cite{ChatRetriever}, a conversationally-tuned retriever built on Qwen-7B~\cite{Qwen}.
For fair comparison, we re-evaluate all baselines under our evaluation setup.

\noindent\textbf{Baseline configurations and query input types.}
We compare different query input strategies. In the rewriting setup, a query rewriter takes the conversation history and current query to produce a rewritten query, which is passed to the retriever. Without rewriting, we consider three variants: (1) \textit{Current}: using only the current query; (2) \textit{Window}: the current query with the last three query-response turns; and (3) \textit{Full}: the entire conversation history concatenated with the current query.

\input{tables/main_table_1}

\subsection{Main Results}
Table~\ref{main_table} presents the comparative evaluation of our approach against existing methods. On TopiOCQA, a benchmark known for its challenging topic shifts within conversations, \myalg{} achieves state-of-the-art performance. On QReCC, our method outperforms most baselines and shows competitive results even against GPT-4-Turbo rewrites.
Without requiring a rewriting process, \myalg{} consistently outperforms GritLM applied to rewritten queries. This indicates that our method effectively leverages rewritten training queries and the inherent generative capability of pre-trained LLMs to contextualize retrieval without needing explicit rewriting at inference time.
Furthermore, our method clearly surpasses ChatRetriever demonstrating that our objective-driven approach to modeling contextual understanding yields more robust performance.

\noindent\textbf{Impact of query input types.}
Experimental results reveal substantial performance differences across query input configurations. The \textit{Current} setting performs poorly due to its inability to resolve abbreviations and lack of contextual cues. Comparisons between \textit{Window} and \textit{Full} configurations highlight key trade-offs: \textit{Window} efficiently captures recent context but may overlook long-range dependencies, while \textit{Full} offers broader coverage at the risk of introducing noise from irrelevant turns. Although dataset-specific input tuning can yield marginal improvements, it lacks generality and relies on heuristic decisions. In contrast, \myalg{} processes the full dialogue holistically and learns to attend to relevant context.

\noindent\textbf{Effectiveness of query rewriting.}
The impact of query rewriting varies across datasets. In TopiOCQA, rewriting consistently improves retrieval performance by resolving context-dependent references and handling topic shifts. LLM-based rewriters such as GritLM and GPT-4-Turbo perform well, effectively capturing nuanced contextual signals. In contrast, T5QR exhibits limited capability, primarily resolving surface-level references such as pronouns (e.g., replacing “it” with its referent). In QReCC, however, rewriting can degrade performance. This degradation is often caused by over-summarization or loss of critical information during rewriting, which removes details necessary for accurate retrieval. In such cases, preserving the original conversational structure proves more effective than rewriting.

\noindent\textbf{Analysis of retriever performance.}
Our analysis indicates that preserving the generative capacity of LLMs plays a crucial role in conversational retrieval. Compared to conventional BERT-based models and LLM-based retrievers trained solely with retrieval objectives (i.e., SFR Embedding), both our model and GritLM incorporate generation loss during training. While all retrievers perform similarly on single-turn queries, regardless of whether the query is original or rewritten, performance gaps widen significantly in multi-turn settings (\textit{Window} and \textit{Full}). This difference is particularly pronounced on TopiOCQA, which involve complex topic shifts.

\input{tables/trec_cast}

\noindent\textbf{Performance on TREC-CAsT.} On the TREC-CAsT benchmark, \myalg{} also delivers strong performance. Notably, our method achieves substantial gains over prior approaches (Table~\ref{trec_cast}). It outperforms recent conversational search methods, including LLM-based query rewriting (e.g., LLM4CS~\cite{LLM4CS}) and conversational retrievers such as ConvDR~\cite{ConvDR}, LeCoRE~\cite{LeCoRE}, ConvAUG~\cite{CONVAUG}, DiSCo~\cite{DiSCo}, and ChatRetriever~\cite{ChatRetriever}. This result demonstrates that even in well-structured evaluation settings, our contextual embedding contributes significantly to improved retrieval quality.

In Appendix~\ref{appendix_ablation} and~\ref{ORConvQA}, further analysis reveals that both our embedding extraction strategy and proposed loss substantially contribute to the performance gains. We also include evaluation results on ORConvQA to confirm the generalizability of our method. Additionally, we report generation performance to validate that our model preserves the language capabilities of the underlying LLM while optimizing for retrieval.

\subsection{Analysis}

\noindent\textbf{Computational cost.}
Table~\ref{computational_cost} summarizes the computational requirements of various query rewriters and retrievers, in terms of model parameters and average inference time. Inference time was measured across 160 samples (10 conversations with 1–16 turns) using an Intel Xeon Gold 6342 CPU and a single NVIDIA RTX A5000 GPU.

\input{tables/computational_cost}

Among query rewriters, we observe substantial differences in both model size and efficiency. All rewriters rely on autoregressive decoding for query generation, which introduces significant latency at inference time. This includes T5QR as well as LLM-based models such as GritLM and GPT-4 Turbo (accessed through OpenAI’s API services). Larger models generally produce higher-quality rewrites, with GPT-4 Turbo delivering the best performance but also incurring the highest cost due to its increased model complexity.

For retrieval, BGE-large achieves the lowest inference time among all evaluated retrievers, reflecting its compact architecture but also its relatively lower retrieval performance compared to larger models. Both \myalg{} and ChatRetriever eliminate the need for separate rewriters and provide end-to-end solutions for conversational search. Notably, \myalg{} further outperforms ChatRetriever in performance, achieving a better balance between effectiveness and efficiency. This supports the advantage of our retriever design in real-world, latency-sensitive applications.

\begin{figure*}[t]
    \centering
    \includegraphics[width=0.97\linewidth]{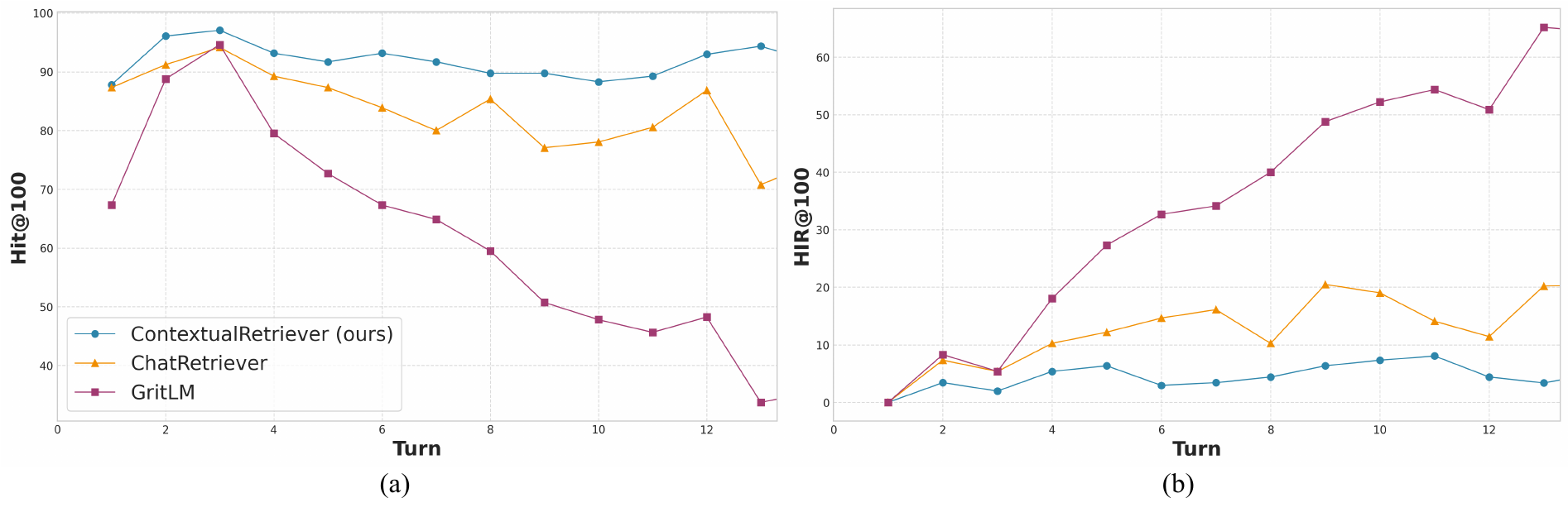}
    \vspace{-3mm}
    \caption{(a) Hit@100 and (b) Historical Interference Rate (HIR@100) of our \myalg{}, ChatRetriever, and GritLM across conversation turns. HIR@100 measures how often a model retrieves passages related to previous queries rather than the current one.}
    \label{fig:analysis3}
\end{figure*}

\noindent\textbf{Ability to capture user intent.}
To evaluate how well models track evolving user intent throughout a dialogue, we conduct a turn-by-turn analysis comparing our method with GritLM and ChatRetriever. This analysis considers not only Hit@k but also whether the retrieved passages reflect the model’s ability to isolate the current information need from earlier conversational turns.

Figure~\ref{fig:analysis3}(a) reports Hit@100 across dialogue turns. Retrieval performance initially improves, as early turns include information that directly supports subsequent queries without significant topic shifts. However, as the conversation progresses, ambiguity and context dependencies accumulate, making retrieval more difficult. GritLM’s performance declines sharply, indicating difficulty in maintaining contextual alignment. ChatRetriever is relatively more stable, while our model consistently achieves higher hit rates, especially in later turns where accurate disambiguation becomes critical.

To further analyze this behavior, we compute Historical Interference Rate (HIR@100), shown in Figure~\ref{fig:analysis3}(b), which measures how often retrieved passages align with ground-truth passages from previous turns rather than the current one. A higher HIR@100 indicates that a model is overly influenced by earlier queries, retrieving outdated or irrelevant content.
GritLM exhibits the highest HIR@100, often retrieving passages aligned with dominant earlier topics regardless of their current relevance. This suggests that the model relies on lexical or shallow semantic cues rather than modeling evolving user intent. ChatRetriever performs better, aided by conversational finetuning, but still suffers from interference. In contrast, our model consistently achieves lower HIR@100 across all turns, demonstrating greater robustness in distinguishing the current query from prior context.

This behavioral distinction is critical. While prior methods may appear context-aware, they often depend on memorization or anchoring to previously relevant contexts. Our model more faithfully tracks shifting user intent, enabling adaptive retrieval even in semantically entangled conversations. These findings reinforce our core design intuition: optimizing for intent-aware representations yields models that are not only accurate but also resilient to context interference and shortcut behaviors. These properties are particularly important in multi-turn settings where user goals evolve continuously.


%% file: tables/ablation1.tex
\begin{table}[t]
\caption{{Retrieval performance comparison:} Baseline models with and without our approach.}
\vspace{-1mm}
\centering
\resizebox{\linewidth}{!}{
\begin{tabular}{l|rrrr}
\toprule
\multirow{2}{*}{Method} &  \multicolumn{2}{c}{TopiOCQA} & \multicolumn{1}{c}{CAsT-19} & \multicolumn{1}{c}{CAsT-20}\\ \cmidrule(lr){2-3} \cmidrule(lr){4-5} 
                     & MRR\textsuperscript{↑} & Hit@100\textsuperscript{↑} & \multicolumn{2}{c}{nDCG@3\textsuperscript{↑}} \\ \midrule
BGE-large  & 16.1 & 46.7 & 36.5 & 20.1 \\
$~~~~$+ ours  & + 7.8 &  + 16.1 & + 13.2 & + 14.8 \\ \midrule
SFR Embedding  & 17.8 & 57.3 & 32.6 & 24.5\\
$~~~~$+ ours  & + 12.4 & + 17.4 & + 8.2 & + 9.6\\ \midrule
GritLM  & 24.3 & 68.5 & 30.7 & 18.2 \\ 
$~~~~$+ ours  & + 17.9 & + 23.4 & + 31.6 & + 28.6 \\ \midrule
\end{tabular}
\label{ablation1}
}
\vspace{-3mm}
\end{table}

%% file: tables/main_table_1.tex
\begin{table*}[t]
\caption{\textbf{Retrieval performance (\%) of different retrievers and query rewriting approaches on TopiOCQA and QReCC datasets.} Best and second-best results are indicated in bold and underlined, respectively. Human rewrites are not included for TopiOCQA as they are not provided in the original dataset.}
\centering
\resizebox{\textwidth}{!}{
\begin{tabular}{c|c|c|>{\centering\arraybackslash}p{1.4cm}>{\centering\arraybackslash}p{1.4cm}>{\centering\arraybackslash}p{1.4cm}|>{\centering\arraybackslash}p{1.4cm}>{\centering\arraybackslash}p{1.4cm}>{\centering\arraybackslash}p{1.4cm}}
\toprule
{\multirow{2}{*}{\raisebox{-0.8ex}{\textbf{Retriever}}}}
& {\multirow{2}{*}{\raisebox{-0.8ex}{\textbf{Query Rewriter}}}}
& {\multirow{2}{*}{\raisebox{-0.8ex}{\textbf{Query Type}}}}
& \multicolumn{3}{c|}{\textbf{TopiOCQA}} 
& \multicolumn{3}{c}{\textbf{QReCC}} \\
\cmidrule(lr){4-6} \cmidrule(lr){7-9}
{} 
& {} 
& {} 
& MRR\textsuperscript{↑} & Hit@20\textsuperscript{↑} & {Hit@100\textsuperscript{↑}}
& MRR\textsuperscript{↑} & Hit@20\textsuperscript{↑} & Hit@100\textsuperscript{↑} \\
\midrule
\multirow{5}{*}{MiniLM} & \multirow{3}{*}{-}  & Current & 3.7  & 9.1 & 13.2  & 4.2 & 10.2 & 14.6 \\
                              &   & Window & 11.1 & 25.8 & 37.5  & 19.7 & 62.1 & 78.1 \\
                              &   & Full & 10.0 & 24.7 & 36.5  & 19.3 & 61.5 & 79.6 \\ \cmidrule{2-9}
                              & GritLM  & \multirow{2}{*}{Rewritten} & 19.8 & 43.6 & 56.5  & 21.5 & 59.9 & 75.7 \\
                              & Human  &  & -  & - & -  & 21.1 & 60.8 & 77.1 \\ \cmidrule{1-9}
                              \multirow{7}{*}{Bge-large} & \multirow{3}{*}{-} & Current & 4.9 & 10.7 & 14.4 & 4.6 & 11.1 & 13.9 \\ 
                               &  & Window & 16.1  & 33.4 & 46.7 & 22.3 & 69.1 & 83.9 \\ 
                               &  & Full & 13.5 & 29.3 & 42.0 & 21.3 & 67.4 & 84.0 \\  \cmidrule{2-9}
                              & T5QR & \multirow{4}{*}{Rewritten} & 18.9 & 42.2 & 55.3 & 16.3 & 49.7 & 64.2 \\ 
                               & GritLM & & 28.3 & 53.0 & 64.1 & 26.0 & 73.7 & 86.2 \\
                              & GPT-4-Turbo & & 37.2 & 71.2 & 82.5 & 28.6 & 80.2 & \textbf{92.0} \\ 
                              & Human &  & - & - & - & 26.3 & 73.9 & 86.8 \\ \midrule
                                     
\multirow{7}{*}{SFR Embedding} & \multirow{3}{*}{-} & Current & 6.1 & 11.3 & 15.0 & 5.3 & 12.8 & 16.1 \\
                             &  & Window & 17.8 & 41.2 & 57.3 & 22.8 & 71.4 & 87.2 \\
                              &  & Full & 14.1 & 31.8 & 46.9 & 21.7 & 68.3 & 86.8 \\ \cmidrule{2-9}
                              & T5QR& \multirow{4}{*}{Rewritten} & 20.6 & 44.8 & 56.6 & 17.7 & 52.7 & 68.4 \\ 
                              & GritLM & & 31.6 & 57.5 & 67.7 & 26.6 & 74.9 & 87.9 \\ 
                              & GPT-4-Turbo & & \underline{40.5} & \underline{76.3} & \underline{86.6} & 28.3 & 79.0 & \underline{91.6} \\ 
                              & Human & & - & - & - & 27.5 & 76.4 & 89.7 \\ \cmidrule{1-9}

\multirow{7}{*}{GritLM} & \multirow{3}{*}{-}  & Current & 2.2  & 10.9 & 11.3  & 3.7 & 9.2  & 12.5 \\
                              &  & Window & 24.3  & 53.4 & 68.5 & 24.9 & 72.5 & 88.2 \\
                              &  & Full & 20.7 & 47.7 & 64.2 & 24.0 & 72.8 & 86.8 \\ \cmidrule{2-9}
                              & T5QR  & \multirow{4}{*}{Rewritten} & 23.6 & 49.3 & 66.2 & 14.7 & 44.4 & 60.6 \\
                              & GritLM  &  & 31.7 & 66.4 & 77.2 & 26.1 & 74.1 & 88.8 \\
                              & GPT-4-Turbo  &  & 35.9 & 69.5 & 83.6 & 26.5 & 74.5 & 88.4 \\
                              & Human &  & - & - & - & 23.2 & 65.6 & 80.7 \\ \cmidrule{1-9}
                              ChatRetriever & - & Full  & {38.1}  & {71.1} & {84.2}  & \underline{36.5} & \underline{82.4} & {91.4} \\ \midrule
                             \textbf{ContextualRetriever (ours)} & - & Full & \textbf{42.2}  & \textbf{81.7} & \textbf{91.9}  & \textbf{36.8} & \textbf{82.7}  & {91.5}\\ \bottomrule
\end{tabular}
\label{main_table}
}
\end{table*}

%% file: tables/trec_cast.tex
\begin{table}[t]
\caption{\textbf{nDCG@3 performance on the TREC-CAsT benchmark.} * indicates the result reported in the original ChatRetriever paper; other results are reproduced. \textbf{+ Response} denotes the use of retrieved responses as additional conversational context.}
\centering
\resizebox{0.8\linewidth}{!}{
\begin{tabular}{lcc}
\toprule
\textbf{Method} & \textbf{CAsT-19} & \textbf{CAsT-20} \\
\midrule
\multicolumn{3}{l}{\textit{Conversational Query Rewriting}} \\
*LLM4CS        & 51.5 & 45.5 \\
\midrule
\multicolumn{3}{l}{\textit{Dense Retrieval}} \\
Bge-large     & 36.5 & 20.1 \\
SFR Embedding & 32.6 & 24.5 \\
GritLM        & 30.7 & 18.2 \\
\midrule
\multicolumn{3}{l}{\textit{Conversational Retrieval}} \\
*ConvDR  & 43.9 & 32.4 \\
*LeCoRE  & 42.2 & 29.0 \\
*ConvAUG  & - & 30.7 \\
*DiSCo (multi-teach)  & - & 35.3 \\
*ChatRetriever & 52.1 & 40.0 \\
ChatRetriever & 54.1 & 38.7 \\
\textbf{ContextualRetriever (ours)} & \textbf{62.3} & \textbf{46.8} \\
$~~~~$\textbf{+ Response} & \textbf{63.4} & \textbf{50.6} \\
\bottomrule
\end{tabular}}
\label{trec_cast}
\vspace{-2mm}
\end{table}

%% file: tables/computational_cost.tex
\begin{table}[]
\caption{Number of parameters (Params.) and inference time for query rewriters and retrievers.}
\centering
\resizebox{\linewidth}{!}{
\begin{tabular}{l|l|l|r}
\toprule
\textbf{Model} & \textbf{Query Type} & \textbf{Params.} & \textbf{Inference Time} \\
\midrule
\multicolumn{4}{l}{\textit{Rewriter}} \\
T5QR & Full & 223M & 156.5ms \\
GritLM & Full & 7241M & 1064.7ms \\
GPT-4-Turbo & Full & -- & 1293.6ms \\
\midrule
\multicolumn{4}{l}{\textit{Retriever}} \\
BGE-large & Window & 335M & 18.4ms \\
GritLM (Current) & Current & 7241M & 35.8ms \\
GritLM (Window) & Window & 7241M & 44.8ms \\
GritLM (Full) & Full & 7241M & 80.3ms \\
ChatRetriever & Full & 7721M & 101.4ms \\
\textbf{ContextualRetriever} & Full & 7241M & 80.5ms \\
\bottomrule
\end{tabular}
\label{computational_cost}
}
\end{table}

%% file: contents/06_conclusion.tex
\section{Conclusions}
We introduced \myalg{}, a unified retriever that generates context-aware embeddings without relying on external query rewriting. Our method integrates user intent understanding directly into the retrieval process by leveraging both conversational context and generation loss during training. Through extensive evaluations on four benchmark datasets, we demonstrate that \myalg{} not only improves retrieval accuracy but also generalizes well across diverse conversational styles and structures. These results suggest that integrating intent modeling within the retriever itself provides a scalable and robust solution for multi-turn conversational search.

\section{Limitations}
Our current implementation fine-tunes base retrievers with LoRA-based parameter-efficient tuning on a multi-turn dataset. While this setup is tailored to multi-turn scenarios and yields strong performance, it may slightly degrade effectiveness on single-turn queries. Extending \myalg{} to a more comprehensive framework that jointly improves both single- and multi-turn performance (e.g., via joint training on combined data) could further enhance generalization. We leave such extensions for future work.


%% file: contents/07_Appendix.tex
\appendix 
\supptitle
\section{Prompt Template for Query Rewriting}
\label{query_rewriting_prompts}
We utilize rewritten queries for intent-guided learning, as described in Eq.~\ref{IGL}. To generate accurate rewritings during training, we leverage both the gold context and the gold response. For each training instance, we employ the GPT-4-Turbo model with the following prompt template.
Note that the prompt includes three few-shot examples, which are manually selected. The rewritten queries in these examples were generated by human annotators.
\begin{tcolorbox}[fonttitle=\small\bfseries, fontupper=\scriptsize\sffamily,
fontlower=\fon{put},
enhanced,
left=2pt, right=2pt, top=0pt, bottom=0pt,
title=Prompt template for query rewriting for training set]
\vspace{-0.5mm}
\begin{lstlisting}[]
Given a previous conversation, a current question,
related context regarding the current question,
and a ground truth response, your task is to
rewrite the current question to make it clearer
and more explicit. In your rewrite, please
avoid using pronouns or any abbreviated terms.
The aim is to ensure that the current question
stands alone, so that LLM can get the related
context and arrive at the ground truth response
without needing additional information.
Do not use any additional comments such as
"Here is a rewritten version of the current
question:". Only generate a rewritten question. 

Examples: {Example1} {Example2} {Example3}

Previous Conversation: {Previous conversation}
Current Question: {Question} 
Context: {Context} 
Ground Truth Response: {Gt_response} 
Output: 
\end{lstlisting}
\end{tcolorbox}

{\noindent}During evaluation, we adopt two prompt-based query rewriters: GritLM and GPT-4-Turbo. Unlike in training, gold context and gold responses are not available at test time. Therefore, rewriting must be performed solely based on the prior conversation history and the current user question.
The prompt template used for GritLM and GPT-4-Turbo during evaluation is provided below:
\begin{tcolorbox}[fonttitle=\small\bfseries, fontupper=\scriptsize\sffamily,
fontlower=\fon{put},
enhanced,
left=2pt, right=2pt, top=0pt, bottom=0pt,
title=Prompt template for query rewriting in evaluation]
\begin{lstlisting}[]
Given a previous conversation and a current
question, your task is to rewrite the current
question to make it clearer and more explicit.
In your rewrite, please avoid using pronouns
or any abbreviated terms. The aim is to ensure
that the current question stands alone, so that
the retriever can get the related context.
If the original question is already clear,
you can use the original question.

Example: {Example1} {Example2} {Example3}

Previous Conversation: {Previous conversation}
Current Question: {Question} 
Rewrite Output:
\end{lstlisting}
\end{tcolorbox}
{\noindent}For GritLM, we enclose the prompt with <|user|> token and add <|assistant|> token before the output.

\section{Evaluation Datasets}
\label{data_stat_section}
We summarize the statistics of the evaluation datasets in Table~\ref{data_statistics}. Our model is trained on the TopiOCQA training split and evaluated on the development sets of all datasets.

\input{tables/data_stats}

\section{Ablation Studies}
All ablation results are reported on a sampled small passage set (5\% of the full dataset), which differs from the main evaluation setup. This allows efficient comparison while preserving relative performance trends.

\label{appendix_ablation}
\subsection{Effect of dialogue history sampling}
As shown in Table~\ref{ablation3}, our dialogue history sampling strategy (Sec.~\ref{sec:sampling}) significantly outperforms the baseline that uses the original training set without augmentation, under the same number of training epochs. These results highlight the effectiveness of our method in exposing the model to diverse conversational contexts and better capturing user intent.
\input{tables/ablation3}

\subsection{Impact of embedding extraction methods}
We analyze our embedding extraction approach described in Sec.~\ref{sec:extraction}, which encodes the full conversation but uses only the current query's embeddings for retrieval. Table~\ref{ablation4} shows that using all output embeddings leads to performance degradation, despite access to full context. This validates our choice to maintain query-focused representations while still leveraging broader dialogue context during encoding.
\input{tables/ablation4}

\subsection{Contribution of learning objectives}
Table~\ref{ablation2} reports the incremental impact of each learning objective. Adding Conversational Contrastive Learning (CCL) to the base configuration substantially improves performance in multi-turn settings. Further gains are observed by incorporating Intent-Guided Learning (IGL) and the Generation Loss (G). These findings confirm that each component contributes to more effective retrieval representations for conversational queries.
\input{tables/ablation2}

\subsection{Hyperparameter senesitivity analysis}
We conducted an ablation study on the weighting coefficients $\lambda_{\text{IGL}}$ and $\lambda_{\text{G}}$ to analyze the sensitivity of our model to these hyperparameters. 
As shown in Table~\ref{tab:lambda_ablation}, the retrieval performance exhibits relatively greater sensitivity to $\lambda_{\text{G}}$, which directly contributes to the retrieval loss. 
In contrast, $\lambda_{\text{IGL}}$ demonstrates more stable trends across different values, indicating that the model is less affected by variations in this parameter.
Overall, the results suggest that careful tuning of $\lambda_{\text{G}}$ is more critical for achieving strong retrieval performance.

\input{tables/ablation5}

\section{Generation Performance}
\label{gen}
Our approach adopts a unified LLM architecture trained with an integrated generation loss, which preserves the model’s ability to generate fluent and contextually relevant responses. To evaluate this capability, we compare our ContextualRetriever with GritLM using the following generation prompt:
\begin{tcolorbox}[fonttitle=\small\bfseries, fontupper=\scriptsize\sffamily,
fontlower=\fon{put},
enhanced,
left=2pt, right=2pt, top=0pt, bottom=0pt,
title=Prompt template for generation]
\begin{lstlisting}[]
<|embed|>\n{Query}\n<|user|>\n {Context}
Optionally using the prior conversation
and context, answer the last query:
{Current Query}\n<|assistant|>\n
\end{lstlisting}
\end{tcolorbox}

\subsection{Evaluation Methodology}
We assess generation performance using gold contexts with two metrics:  
(1) \textbf{Lexical Matching}~\cite{GenMetric1, GenMetric2}, which measures whether the generated answer contains any reference answer span; and (2) \textbf{Correctness}~\cite{zhong2024memorybank}, which uses GPT-4-Turbo to score semantic correctness with the following evaluation prompt:

\begin{tcolorbox}[fonttitle=\small\bfseries, fontupper=\scriptsize\sffamily,
fontlower=\fon{put},
enhanced,
left=2pt, right=2pt, top=0pt, bottom=0pt,
title=Prompt template for measuring correctness]
\begin{lstlisting}[]
Evaluates if the response contains the correct
answer to the probing question
(labels: (0 : wrong, 0.5 : partial, 1 : correct)
You should return only digit 0, 0.5, or 1.
response: {response} answer: {answer}
\end{lstlisting}
\end{tcolorbox}

\subsection{Results and Analysis}
As shown in Table~\ref{gen_performance}, our model achieves superior generation accuracy compared to GritLM, validating the effectiveness of our approach in maintaining robust generation capabilities.

\input{tables/generation_performance}

{\noindent}Notably, our unified architecture enables efficient cache sharing between retrieval and generation. During inference, the query representations computed for retrieval can be reused for generation without incurring additional computational cost.

\input{tables/orconvqa}
\section{Results on ORConvQA}
\label{ORConvQA}
We evaluate our model against various baselines on the ORConvQA dataset, which is characterized by consistent topic maintenance without shifts. Our experiments reveal that \textit{Full} setting achieves the best performance due to the dataset's preference for comprehensive information preservation, while query rewriting approaches showed lower performance due to information loss. Notably, the BERT-based retriever (Bge-large) demonstrates strong performance on this dataset, primarily due to the strong correlation between previous conversations and current queries, and limited requirement for complex user intent understanding. Despite GritLM's relatively lower baseline performance, our approach achieves state-of-the-art results, outperforming both existing baselines and ChatRetriever across different query configurations.

%% file: tables/data_stats.tex
\begin{table}[h]
\caption{Statistics of the datasets, including the number of conversations (C), queries (Q), and passages (P).}
\centering
\resizebox{1.0\linewidth}{!}{
\begin{tabular}{c|cc|c|cc|c}
\toprule
\multirow{2}{*}{Statistics} &  \multicolumn{2}{c|}{TopiOCQA} & QReCC & CAsT-19 & CAsT-20 & ORConvQA \\ \cline{2-7}
                            & Train & Test & Dev & Test & Test & Test \\ \midrule
C &  3,509 & 205 & 2,000 & 50 & 25 & 490 \\ 
Q & 45,450 & 2,514 & 11,573 & 479 & 208 & 3,430 \\
P & \multicolumn{2}{c|}{25M} & {54M} & \multicolumn{2}{c|}{38M} & {11M} \\ \bottomrule
\end{tabular}
\label{data_statistics}
}
\end{table}

%% file: tables/ablation3.tex
\begin{table}[h]
\caption{Performance comparison (\%) between baseline and our sampling strategy.}
\centering
\resizebox{0.88\linewidth}{!}{
\begin{tabular}{c|ccc}
\toprule
\multirow{2}{*}{Sampling strategy} &  \multicolumn{2}{c}{TopiOCQA} \\ \cline{2-3} 
                     & nDCG@3 & Hit@5 \\ \midrule
Baseline & 39.3 & 80.9 \\
Dialogue history sampling (ours) & 45.5 & 88.3 \\ \midrule
\end{tabular}
\label{ablation3}
}
\end{table}

%% file: tables/ablation4.tex
\begin{table}[h]
\caption{Performance comparison (\%) of different retrieval embedding extraction methods.}
\centering
\resizebox{\linewidth}{!}{
\begin{tabular}{c|c|ccc}
\toprule
\multirow{2}{*}{Retrieval Embeddings} & \multirow{2}{*}{Retriever} & \multicolumn{2}{c}{TopiOCQA} \\ \cline{3-4} 
                    & & nDCG@3 & Hit@5 \\ \midrule
\multirow{1}{*}{Full conversation}  
                                    & GritLM & 15.9 & 41.5 \\ \midrule
\multirow{2}{*}{Current query-focused} 
                                    & GritLM & 29.2 & 64.0 \\
                                    & ContextualRetriever & 45.5 & 88.3  \\ \midrule
\end{tabular}
\label{ablation4}
}
\end{table}

%% file: tables/ablation2.tex
\begin{table}[h]
\caption{Impact of different learning components.}
\centering
\resizebox{0.8\linewidth}{!}{
\begin{tabular}{c|ccc}
\toprule
\multirow{2}{*}{Method} &  \multicolumn{2}{c}{TopiOCQA} \\ \cline{2-3} 
                     & nDCG@3 & Hit@5 \\ \midrule
GritLM & 23.3  & 58.0 \\ \midrule
+ $L_{CCL}$ & 40.8  & 79.0 \\
+ $L_{CCL} + L_{IGL}$  & 42.0 & 83.5 \\
+ $L_{CCL} + L_{IGL} + L_{G}$ & 45.5 & 88.3 \\ \midrule
\end{tabular}
\label{ablation2}
}
\end{table}

%% file: tables/ablation5.tex
\begin{table}[h]
\caption{Ablation on $\lambda_{\text{IGL}}$ and $\lambda_{\text{G}}$ on TopiOCQA.}
\centering
\resizebox{0.6\linewidth}{!}{
\begin{tabular}{c|c|cc}
\toprule
$\lambda_{\text{IGL}}$ & $\lambda_{\text{G}}$ & \multicolumn{2}{c}{TopiOCQA} \\
\cline{3-4}
 & & nDCG@3 & Hit@5 \\ \midrule
1.0 & 0.05 & 43.2 & 86.2 \\
1.0 & 0.10 & 45.5 & 88.3 \\
1.0 & 0.30 & 43.8 & 87.0 \\
0.5 & 0.10 & 45.0 & 88.1 \\
2.0 & 0.10 & 45.2 & 89.1 \\ \midrule
\end{tabular}
\label{tab:lambda_ablation}
}
\end{table}

%% file: tables/generation_performance.tex
\begin{table}[h]
\caption{Generation performance of Unified LLM.}
\centering
\resizebox{0.9\linewidth}{!}{
\begin{tabular}{c|cc}
\toprule
\multirow{2}{*}{Model} &  \multicolumn{2}{c}{Gen. performance (\%)} \\ \cline{2-3}
                        &  Lexical Matching & Correctness \\ \midrule
{GritLM} & 27.6 & 60.5 \\
{\textbf{ContextualRetriever}} & 31.7 & 70.3 \\ \bottomrule
\end{tabular}
\label{gen_performance}
}
\end{table}

%% file: tables/orconvqa.tex
\begin{table*}[t]
\caption{Retrieval performance (\%) of different retrievers and query rewriting approaches on ORConvQA.}
\centering
\resizebox{0.6\textwidth}{!}{
\begin{tabular}{c|c|c|cc}
\toprule
\multirow{2}{*}{Retriever} & \multirow{2}{*}{Query Rewriter} &  \multirow{2}{*}{Query Type} & \multicolumn{2}{c}{ORConvQA} \\ \cline{4-5} 
& & & nDCG@3 & Hit@5 \\ \midrule
\multirow{4}{*}{MiniLM} & \multirow{3}{*}{-}  & Current & 8.2 & 16.2 \\
                            &   & Window &  46.5 & 74.1  \\
                            & & Full & 62.5  &  90.0 \\ \cmidrule{2-5}
                            & GritLM  & Rewritten & 40.6  & 69.0 \\ \cmidrule{1-5}
\multirow{7}{*}{Bge-large} & \multirow{3}{*}{-} & Current & 12.3 & 20.6 \\ 
                             &  & Window & 60.2  & 88.5 \\ 
                             &  & Full &  71.4 & \underline{97.0} \\ \cmidrule{2-5}
                            & T5QR & \multirow{4}{*}{Rewritten} & 44.0 & 68.8 \\ 
                             & GritLM & & 56.0 & 84.3  \\
                            & GPT-4 & &  65.4 & 94.2 \\ 
                            & Human &  & 54.3 & 82.9  \\ \midrule
                                     
\multirow{7}{*}{SFR Embedding} & \multirow{3}{*}{-} & Current & 11.7 &  20.1  \\
                            &  & Window & 62.7 & 88.9 \\
                            & & Full & 72.9 & 94.4 \\ \cmidrule{2-5}
                            & T5QR& \multirow{4}{*}{Rewritten} & 40.0 & 65.5 \\ 
                            & GritLM & & 48.9 & 77.8 \\ 
                            & GPT-4 & & 59.0 & 89.3 \\ 
                            & Human & & 50.5 & 79.7  \\ \cmidrule{1-5}

\multirow{7}{*}{GritLM} & \multirow{3}{*}{-}  & Current & 10.8  &  17.4 \\
                            &  & Window & 59.1 & 84.4 \\
                            &  & Full &  \textbf{73.1}  & 95.2 \\ \cmidrule{2-5}
                            & T5QR  & \multirow{4}{*}{Rewritten} & 36.9  & 59.2  \\
                            & GritLM  &  & 46.1  & 71.7  \\
                            & GPT-4  &  & 55.5 & 82.8\\ 
                            & Human &  & 45.5 & 71.1  \\ \cmidrule{1-5}
                             ChatRetriever & - & Full  &   70.6 & 95.0 \\ \midrule
                            \textbf{ContextualRetriever (ours)} & - & Full & \underline{73.0}  & \textbf{97.8}  \\ \midrule
\end{tabular}
\label{orconvqa}
}
\end{table*}

%% file: main.bbl
\begin{thebibliography}{40}
\expandafter\ifx\csname natexlab\endcsname\relax\def\natexlab#1{#1}\fi

\bibitem[{Achiam et~al.(2023)Achiam, Adler, Agarwal, Ahmad, Akkaya, Aleman, Almeida, Altenschmidt, Altman, Anadkat et~al.}]{GPT4}
Josh Achiam, Steven Adler, Sandhini Agarwal, Lama Ahmad, Ilge Akkaya, Florencia~Leoni Aleman, Diogo Almeida, Janko Altenschmidt, Sam Altman, Shyamal Anadkat, et~al. 2023.
\newblock Gpt-4 technical report.
\newblock \emph{arXiv preprint arXiv:2303.08774}.

\bibitem[{Adlakha et~al.(2022)Adlakha, Dhuliawala, Suleman, de~Vries, and Reddy}]{TopiOCQA}
Vaibhav Adlakha, Shehzaad Dhuliawala, Kaheer Suleman, Harm de~Vries, and Siva Reddy. 2022.
\newblock \href {https://doi.org/10.1162/tacl_a_00471} {Topi{OCQA}: Open-domain conversational question answering with topic switching}.
\newblock \emph{Transactions of the Association for Computational Linguistics}, 10:468--483.

\bibitem[{Anantha et~al.(2021)Anantha, Vakulenko, Tu, Longpre, Pulman, and Chappidi}]{QReCC}
Raviteja Anantha, Svitlana Vakulenko, Zhucheng Tu, Shayne Longpre, Stephen Pulman, and Srinivas Chappidi. 2021.
\newblock Open-domain question answering goes conversational via question rewriting.
\newblock \emph{Proceedings of the 2021 Conference of the North American Chapter of the Association for Computational Linguistics: Human Language Technologies}.

\bibitem[{Bai et~al.(2023)Bai, Bai, Chu, Cui, Dang, Deng, Fan, Ge, Han, Huang et~al.}]{Qwen}
Jinze Bai, Shuai Bai, Yunfei Chu, Zeyu Cui, Kai Dang, Xiaodong Deng, Yang Fan, Wenbin Ge, Yu~Han, Fei Huang, et~al. 2023.
\newblock Qwen technical report.
\newblock \emph{arXiv preprint arXiv:2309.16609}.

\bibitem[{Chen et~al.(2024)Chen, Dou, Mao, Liu, and Zhao}]{CONVAUG}
Haonan Chen, Zhicheng Dou, Kelong Mao, Jiongnan Liu, and Ziliang Zhao. 2024.
\newblock Generalizing conversational dense retrieval via llm-cognition data augmentation.
\newblock \emph{arXiv preprint arXiv:2402.07092}.

\bibitem[{Dalton et~al.(2020)Dalton, Xiong, and Callan}]{TREC19}
Jeffrey Dalton, Chenyan Xiong, and Jamie Callan. 2020.
\newblock Trec cast 2019: The conversational assistance track overview.
\newblock In \emph{In Proceedings of TREC}.

\bibitem[{Dalton et~al.(2021)Dalton, Xiong, and Callan}]{TREC20}
Jeffrey Dalton, Chenyan Xiong, and Jamie Callan. 2021.
\newblock Cast 2020: The conversational assistance track overview.
\newblock In \emph{In Proceedings of TREC}.

\bibitem[{Gao et~al.(2023)Gao, Xiong, Bennett, and Craswell}]{ConvSearch1}
Jianfeng Gao, Chenyan Xiong, Paul Bennett, and Nick Craswell. 2023.
\newblock \emph{Neural approaches to conversational information retrieval}.
\newblock Springer.

\bibitem[{Izacard and Grave(2021)}]{GenMetric2}
Gautier Izacard and Edouard Grave. 2021.
\newblock Leveraging passage retrieval with generative models for open domain question answering.
\newblock In \emph{Proceedings of the 16th Conference of the European Chapter of the Association for Computational Linguistics: Main Volume}, pages 874--880. Association for Computational Linguistics.

\bibitem[{Jiang et~al.(2023)Jiang, Sablayrolles, Mensch, Bamford, Chaplot, Casas, Bressand, Lengyel, Lample, Saulnier et~al.}]{Mistral}
Albert~Q Jiang, Alexandre Sablayrolles, Arthur Mensch, Chris Bamford, Devendra~Singh Chaplot, Diego de~las Casas, Florian Bressand, Gianna Lengyel, Guillaume Lample, Lucile Saulnier, et~al. 2023.
\newblock Mistral 7b.
\newblock \emph{arXiv preprint arXiv:2310.06825}.

\bibitem[{Karpukhin et~al.(2020)Karpukhin, O{\u{g}}uz, Min, Lewis, Wu, Edunov, Chen, and Yih}]{DPR}
Vladimir Karpukhin, Barlas O{\u{g}}uz, Sewon Min, Patrick Lewis, Ledell Wu, Sergey Edunov, Danqi Chen, and Wen-tau Yih. 2020.
\newblock Dense passage retrieval for open-domain question answering.
\newblock \emph{arXiv preprint arXiv:2004.04906}.

\bibitem[{Kenton and Toutanova(2019)}]{BERT}
Jacob Devlin Ming-Wei~Chang Kenton and Lee~Kristina Toutanova. 2019.
\newblock Bert: Pre-training of deep bidirectional transformers for language understanding.
\newblock In \emph{Proceedings of naacL-HLT}, volume~1, page~2. Minneapolis, Minnesota.

\bibitem[{Khattab and Zaharia(2020)}]{ColBERT}
Omar Khattab and Matei Zaharia. 2020.
\newblock Colbert: Efficient and effective passage search via contextualized late interaction over bert.
\newblock In \emph{Proceedings of the 43rd International ACM SIGIR conference on research and development in Information Retrieval}, pages 39--48.

\bibitem[{Kim and Kim(2022)}]{ShortCut}
Sungdong Kim and Gangwoo Kim. 2022.
\newblock Saving dense retriever from shortcut dependency in conversational search.
\newblock \emph{arXiv preprint arXiv:2202.07280}.

\bibitem[{Lee et~al.(2024)Lee, Roy, Xu, Raiman, Shoeybi, Catanzaro, and Ping}]{NV1}
Chankyu Lee, Rajarshi Roy, Mengyao Xu, Jonathan Raiman, Mohammad Shoeybi, Bryan Catanzaro, and Wei Ping. 2024.
\newblock Nv-embed: Improved techniques for training llms as generalist embedding models.
\newblock \emph{arXiv preprint arXiv:2405.17428}.

\bibitem[{Li et~al.(2024)Li, Qin, Xiao, Chen, Luo, Shao, Lian, and Liu}]{BGE-ICL}
Chaofan Li, MingHao Qin, Shitao Xiao, Jianlyu Chen, Kun Luo, Yingxia Shao, Defu Lian, and Zheng Liu. 2024.
\newblock \href {http://arxiv.org/abs/2409.15700} {Making text embedders few-shot learners}.

\bibitem[{Lin et~al.(2021)Lin, Yang, and Lin}]{CQE}
Sheng-Chieh Lin, Jheng-Hong Yang, and Jimmy Lin. 2021.
\newblock Contextualized query embeddings for conversational search.
\newblock In \emph{Findings of the Association for Computational Linguistics: EMNLP 2023}.

\bibitem[{Lin et~al.(2020)Lin, Yang, Nogueira, Tsai, Wang, and Lin}]{T5QR}
Sheng-Chieh Lin, Jheng-Hong Yang, Rodrigo Nogueira, Ming-Feng Tsai, Chuan-Ju Wang, and Jimmy Lin. 2020.
\newblock Conversational question reformulation via sequence-to-sequence architectures and pretrained language models.
\newblock \emph{arXiv preprint arXiv:2004.01909}.

\bibitem[{Lupart et~al.(2024)Lupart, Aliannejadi, and Kanoulas}]{DiSCo}
Simon Lupart, Mohammad Aliannejadi, and Evangelos Kanoulas. 2024.
\newblock Disco meets llms: A unified approach for sparse retrieval and contextual distillation in conversational search.
\newblock \emph{arXiv preprint arXiv:2410.14609}.

\bibitem[{Mao et~al.(2024)Mao, Deng, Chen, Mo, Liu, Sakai, and Dou}]{ChatRetriever}
Kelong Mao, Chenlong Deng, Haonan Chen, Fengran Mo, Zheng Liu, Tetsuya Sakai, and Zhicheng Dou. 2024.
\newblock Chatretriever: Adapting large language models for generalized and robust conversational dense retrieval.
\newblock \emph{Proceedings of the 2024 Conference on Empirical Methods in Natural Language Processing}.

\bibitem[{Mao et~al.(2023{\natexlab{a}})Mao, Dou, Mo, Hou, Chen, and Qian}]{LLM4CS}
Kelong Mao, Zhicheng Dou, Fengran Mo, Jiewen Hou, Haonan Chen, and Hongjin Qian. 2023{\natexlab{a}}.
\newblock Large language models know your contextual search intent: A prompting framework for conversational search.
\newblock \emph{arXiv preprint arXiv:2303.06573}.

\bibitem[{Mao et~al.(2023{\natexlab{b}})Mao, Qian, Mo, Dou, Liu, Cheng, and Cao}]{LeCoRE}
Kelong Mao, Hongjin Qian, Fengran Mo, Zhicheng Dou, Bang Liu, Xiaohua Cheng, and Zhao Cao. 2023{\natexlab{b}}.
\newblock Learning denoised and interpretable session representation for conversational search.
\newblock In \emph{Proceedings of the ACM Web Conference 2023}.

\bibitem[{Meng et~al.(2024)Meng, Liu, Joty, Xiong, Zhou, and Yavuz}]{SFR}
Rui Meng, Ye~Liu, Shafiq~Rayhan Joty, Caiming Xiong, Yingbo Zhou, and Semih Yavuz. 2024.
\newblock Sfrembedding-mistral: enhance text retrieval with transfer learning.
\newblock \emph{Salesforce AI Research Blog}, 3.

\bibitem[{Mo et~al.(2024)Mo, Mao, Zhao, Qian, Chen, Cheng, Li, Zhu, Dou, and Nie}]{ConvSearch3}
Fengran Mo, Kelong Mao, Ziliang Zhao, Hongjin Qian, Haonan Chen, Yiruo Cheng, Xiaoxi Li, Yutao Zhu, Zhicheng Dou, and Jian-Yun Nie. 2024.
\newblock A survey of conversational search.
\newblock \emph{arXiv preprint arXiv:2410.15576}.

\bibitem[{Mo et~al.(2023)Mo, Mao, Zhu, Wu, Huang, and Nie}]{ConvGQR}
Fengran Mo, Kelong Mao, Yutao Zhu, Yihong Wu, Kaiyu Huang, and Jian-Yun Nie. 2023.
\newblock Convgqr: generative query reformulation for conversational search.
\newblock \emph{Association for Computational Linguistics}.

\bibitem[{Moreira et~al.(2024)Moreira, Osmulski, Xu, Ak, Schifferer, and Oldridge}]{NV2}
Gabriel de Souza~P Moreira, Radek Osmulski, Mengyao Xu, Ronay Ak, Benedikt Schifferer, and Even Oldridge. 2024.
\newblock Nv-retriever: Improving text embedding models with effective hard-negative mining.
\newblock \emph{arXiv preprint arXiv:2407.15831}.

\bibitem[{Muennighoff et~al.(2024)Muennighoff, Su, Wang, Yang, Wei, Yu, Singh, and Kiela}]{GritLM}
Niklas Muennighoff, Hongjin Su, Liang Wang, Nan Yang, Furu Wei, Tao Yu, Amanpreet Singh, and Douwe Kiela. 2024.
\newblock Generative representational instruction tuning.
\newblock \emph{arXiv preprint arXiv:2402.09906}.

\bibitem[{Muennighoff et~al.(2022)Muennighoff, Tazi, Magne, and Reimers}]{MTEB}
Niklas Muennighoff, Nouamane Tazi, Lo{\"\i}c Magne, and Nils Reimers. 2022.
\newblock Mteb: Massive text embedding benchmark.
\newblock \emph{arXiv preprint arXiv:2210.07316}.

\bibitem[{Qu et~al.(2020)Qu, Yang, Chen, Qiu, Croft, and Iyyer}]{ORConvQA}
Chen Qu, Liu Yang, Cen Chen, Minghui Qiu, W.~Bruce Croft, and Mohit Iyyer. 2020.
\newblock {Open-Retrieval Conversational Question Answering}.
\newblock In \emph{SIGIR}.

\bibitem[{Raffel et~al.(2020)Raffel, Shazeer, Roberts, Lee, Narang, Matena, Zhou, Li, and Liu}]{t5base}
Colin Raffel, Noam Shazeer, Adam Roberts, Katherine Lee, Sharan Narang, Michael Matena, Yanqi Zhou, Wei Li, and Peter~J. Liu. 2020.
\newblock \href {http://jmlr.org/papers/v21/20-074.html} {Exploring the limits of transfer learning with a unified text-to-text transformer}.
\newblock \emph{Journal of Machine Learning Research}, 21(140):1--67.

\bibitem[{Ramos et~al.(2003)}]{TF-IDF}
Juan Ramos et~al. 2003.
\newblock Using tf-idf to determine word relevance in document queries.
\newblock In \emph{Proceedings of the first instructional conference on machine learning}, volume 242, pages 29--48. Citeseer.

\bibitem[{Robertson et~al.(2009)Robertson, Zaragoza et~al.}]{BM25}
Stephen Robertson, Hugo Zaragoza, et~al. 2009.
\newblock The probabilistic relevance framework: Bm25 and beyond.
\newblock \emph{Foundations and Trends{\textregistered} in Information Retrieval}, 3(4):333--389.

\bibitem[{Touvron et~al.(2023)Touvron, Lavril, Izacard, Martinet, Lachaux, Lacroix, Rozi{\`e}re, Goyal, Hambro, Azhar et~al.}]{Llama}
Hugo Touvron, Thibaut Lavril, Gautier Izacard, Xavier Martinet, Marie-Anne Lachaux, Timoth{\'e}e Lacroix, Baptiste Rozi{\`e}re, Naman Goyal, Eric Hambro, Faisal Azhar, et~al. 2023.
\newblock Llama: Open and efficient foundation language models.
\newblock \emph{arXiv preprint arXiv:2302.13971}.

\bibitem[{Wang et~al.(2023)Wang, Cheng, Guo, Yue, Ding, Xu, Wang, Hu, Zhang, and Zhang}]{GenMetric1}
Cunxiang Wang, Sirui Cheng, Qipeng Guo, Yuanhao Yue, Bowen Ding, Zhikun Xu, Yidong Wang, Xiangkun Hu, Zheng Zhang, and Yue Zhang. 2023.
\newblock \href {https://openreview.net/forum?id=UErNpveP6R} {Evaluating open-{QA} evaluation}.
\newblock In \emph{Thirty-seventh Conference on Neural Information Processing Systems Datasets and Benchmarks Track}.

\bibitem[{Wang et~al.(2020)Wang, Wei, Dong, Bao, Yang, and Zhou}]{MiniLM}
Wenhui Wang, Furu Wei, Li~Dong, Hangbo Bao, Nan Yang, and Ming Zhou. 2020.
\newblock Minilm: Deep self-attention distillation for task-agnostic compression of pre-trained transformers.
\newblock \emph{Advances in Neural Information Processing Systems}, 33:5776--5788.

\bibitem[{Xiao et~al.(2023)Xiao, Liu, Zhang, and Muennighof}]{BGE}
Shitao Xiao, Zheng Liu, Peitian Zhang, and Niklas Muennighof. 2023.
\newblock C-pack: Packaged resources to advance general chinese embedding.
\newblock \emph{arXiv preprint arXiv:2309.07597}.

\bibitem[{Ye et~al.(2023)Ye, Fang, Li, and Yilmaz}]{ConversatioanlReWrite}
Fanghua Ye, Meng Fang, Shenghui Li, and Emine Yilmaz. 2023.
\newblock \href {https://doi.org/10.18653/v1/2023.findings-emnlp.398} {Enhancing conversational search: Large language model-aided informative query rewriting}.
\newblock In \emph{Findings of the Association for Computational Linguistics: EMNLP 2023}, pages 5985--6006, Singapore. Association for Computational Linguistics.

\bibitem[{Yu et~al.(2021)Yu, Liu, Xiong, Feng, and Liu}]{ConvDR}
Shi Yu, Zhenghao Liu, Chenyan Xiong, Tao Feng, and Zhiyuan Liu. 2021.
\newblock Few-shot conversational dense retrieval.
\newblock In \emph{Proceedings of the 44th International ACM SIGIR Conference on research and development in information retrieval}, pages 829--838.

\bibitem[{Zhang et~al.(2024)Zhang, Peng, Sun, Chen, Liang, Zhang, Zhou, Chen, and Zhang}]{OneGen}
Jintian Zhang, Cheng Peng, Mengshu Sun, Xiang Chen, Lei Liang, Zhiqiang Zhang, Jun Zhou, Huajun Chen, and Ningyu Zhang. 2024.
\newblock Onegen: Efficient one-pass unified generation and retrieval for llms.
\newblock In \emph{Findings of the Association for Computational Linguistics: EMNLP 2024}, pages 4088--4119.

\bibitem[{Zhong et~al.(2024)Zhong, Guo, Gao, Ye, and Wang}]{zhong2024memorybank}
Wanjun Zhong, Lianghong Guo, Qiqi Gao, He~Ye, and Yanlin Wang. 2024.
\newblock Memorybank: Enhancing large language models with long-term memory.
\newblock In \emph{Proceedings of the AAAI Conference on Artificial Intelligence}, volume~38, pages 19724--19731.

\end{thebibliography}
